# Band Alignment in Molecular Devices: Influence of Anchoring Group and Metal Work Function

Jian-guo Wang[1], Emil Prodan[2*], Roberto Car[1], Annabella Selloni[1*]

[1]Department of Chemistry, Princeton University, Princeton, NJ-08540, USA.

[2]Department of Physics, Yeshiva University, New York, NY 10016, USA.

E-mail: prodan@yu.edu, aselloni@princeton.edu

**We present periodic Density Functional Theory calculations of the electronic properties of molecular junctions formed by amine-, and thiol-terminated alkane chains attached to two metal (Au, Ag) electrodes. Based on extensive analysis that includes molecular monolayers of varying densities, we establish a relationship between the alignment of the molecular energy levels and the interface dipoles, which shows that the band alignment (BA) in the limit of long, isolated chains is independent of the link group and can be computed from a reference system of non interacting molecule + metal electrodes. The main difference between the amine and thiol linkers is the effective dipole moment at the contact. This is very large, about 4.5 D, for amine linkers, leading to a strong dependence of the BA on the monolayer density and a slow convergence to the isolated molecule limit. Instead, this convergence is fast for S anchors due to the very small, ~ 0.2 D, effective dipoles at the contacts.**

The linear, small bias potential, tunneling transport through insulating molecular chains is strongly dependent on the alignment of the energy levels of the molecular chain relative to the Fermi energy of the (infinite) metallic electrodes. More precisely, the band alignment (BA) determines the coefficient $\beta$ in the exponential variation of the conductance $g$ with $n$, the number of monomers in the chain: $g = g_c e^{-\beta n}$.[1-8] For this reason, a large body of experimental and theoretical work has been devoted to understanding the BA in metal-molecule-metal junctions and how this is related to the character (e.g. aromatic or aliphatic) of the molecular chains, to the chemical bond (or contact) between the molecule and the electrode, and to the properties of the electrode material (e.g. the metal work function).[9-14] A great deal of attention has been focused on devices formed by thiol-terminated molecules and gold electrodes, where the contact is established through strong (~ 2 eV) S-Au bonds.[1, 9, 15-18] Recently, however, interest in alternative chemical linker groups has emerged.[19-23] For the particular case of alkane chains, experiments have shown that while the chemical nature of the linker has a weak influence on $\beta$, it has a rather strong effect on the prefactor $g_c$ and thus on the overall current.[22, 24] Similarly, the electrode work function was found to affect only weakly $\beta$.[7] A few theoretical investigations have already attempted to rationalize these observations,[21,25] but the specific effect of different linker groups on the BA is not well understood yet.

In this paper, we present periodic Density Functional Theory (DFT) calculations of the electronic properties of prototypical M-X(CH$_2$)$_n$X-M devices formed by alkane chains (CH$_2$)$_n$ chemically bonded to two metal (M) electrodes via different anchoring groups (X). We mainly focus on Au(111) electrodes and amine (X = NH$_2$) linkers, which have recently attracted considerable interest, but, for comparison, we also examine the "classic" thiolate (X=S) anchoring groups as well as Ag(111) electrodes. To investigate the effect of the chain's length, we consider alkane chains with n = 6 and n = 16. As a consequence of the adopted periodic boundary conditions, molecular monolayers rather than isolated molecules are actually simulated, and the dependence on the monolayer density is studied. We investigate the alignment of the molecular levels with the Fermi energy, $E_F$, of the electrodes and how this is related to the electrostatic potentials and the interface dipoles arising from the charge redistribution that follows the formation of a chemical bond between the metal and the anchoring group. Based on this comprehensive analysis, we establish a relationship between the BA and the interface dipoles that accounts for the role of the anchoring group and of the electrode work function, $\Phi$, on the electronic properties of the device.

A simple sketch of our molecular device is shown in Fig. 1a, where some of the properties of interest to the present study are also indicated. When studying electrostatics and BA, it is useful to introduce a reference system consisting of the non-interacting metal electrodes and the molecular monolayer. The latter includes the alkane chains and the linker groups in the same geometry of the actual device. For example, we can imagine that the reference system results from a continuous process in which the left/right Au atoms are rigidly shifted to the left/right, while keeping the geometry of the monolayer (including the linkers) fixed, as schematically illustrated in Fig. 1b for Au-NH$_2$(CH$_2$)$_6$NH$_2$-Au. After this process is completed, electric dipoles appear at the metal and monolayer surfaces, as schematically indicated in Fig. 1. We denote the corresponding dipoles per surface unit cell by $p_{\text{metal}}$ and $p_{\text{mol}}$, and denote by $p_{\text{device}}$ the dipole present at the contact of the device before performing the process that led to the reference system. In Fig. 1, the arrows correspond to the positive orientation of the dipoles, but the latter can also take negative values. A precise definition of the dipole moments will be given later.

We carried out the DFT calculations at the generalized gradient approximation (GGA) level, using the PW91 functional,[26] within a plane wave-pseudopotential scheme. Computational details are given in the Supporting Information (SI). The molecular junctions were modeled using a repeated slab geometry, in which the molecules are sandwiched between Au(111) slabs made of four layers. Supercells of different lateral sizes were considered: (p × p) supercells with p=2, 3, 4, and 6 – containing $n_L = p^2$ Au atoms per layer -- in the case of $NH_2$-terminated molecules, and $(q\sqrt{3} \times q\sqrt{3})R30$ with q=1, 2 and 3 -- containing $n_L = 3q^2$ Au atoms per layer -- for alkanedithiols. The smallest p/q value corresponds to a dense monolayer, while the largest p/q value corresponds to a low coverage situation, i.e. as close as we can get to the isolated molecule limit. With this setup, we find $\Phi$ = 4.5 ± 0.1 eV (expt:4.74 eV[27]) and $\Phi$ = 5.3 ± 0.1 eV (expt: 5.31 eV[27]) for the clean Ag(111) and Au(111) surfaces, respectively.

*Geometries* - In agreement with previous theoretical studies,[28] the adsorption of the $NH_2$-terminated molecules is found to be very weak on the defect free Au(111) surface. A somewhat stronger binding (Ea= 0.66 eV for $NH_2(CH_2)_6NH_2$) is obtained for adsorption on a Au adatom, a geometry that is likely to occur at break junctions.[19-21] The latter type of geometry is thus used to study the junctions with diamine molecules (top left panel of Fig. 2). In the optimized structure for Au-$NH_2C_6H_{12}NH_2$-Au, the distance between the nitrogen headgroup and the Au adatom (at the surface fcc hollow site) is 2.375 Å, with a Au-N-C angle of ~ 124°. For Au-$SC_nH_{2n}S$-Au junctions, the metal-molecule contact geometry was derived from the one theoretically predicted for the ($\sqrt{3} \times \sqrt{3}$) high density phase of alkanethiol Self Assembled Monolayers on Au(111):[29] both S headgroups are directly adsorbed at a bridge-fcc site of the Au(111) electrode (without extra Au adatoms), and the alkane chains are tilted by ~ 25° (top right panel of Fig. 2).

*Electrostatics.* We examine the change in the charge density, $\Delta n(\mathbf{r})$, and in the electrostatic potential energy, $\Delta V(\mathbf{r})$, between a device and the corresponding reference system, focusing on the dependence upon the lateral size of the supercell (Fig. 2). This dependence is important for investigating the isolated chain limit, which can be achieved by increasing the supercell area, $A$. $\Delta n(\mathbf{r})$ shows the charge redistribution occurring at the contact between a metallic electrode and a molecule. The resulting interface dipole affects the electrostatic potential at the interface, as illustrated in Fig. 2, which reports the planar ($x,y$) average of $\Delta V(\mathbf{r})$ for Au(111) and Ag(111) metal electrodes and for both linker groups X= $NH_2$ and X= S. In this figure, the different curves refer to calculations with different (p × p) supercells, corresponding to different molecular coverages. The main feature in Fig. 2 is the strong dependence of $\Delta V(z)$ on coverage.

For $NH_2$ anchors, $\Delta V$ drops sharply at the contacts and becomes approximately constant inside the molecule. The potential drop – larger for Au electrodes - decreases with increasing $A$, with a behavior that becomes approximately linear in ($1/A$) when p ≥ 3 (see Fig. 4), as one would expect for a localized dipole at the contact with the electrode. The potential drop suggests a charge transfer from the molecule to the metal, consistent with the electron-donor character of the amine linkers.

The behavior of $\Delta V$ is very different in the case of thiol-terminated chains. First, the absolute value of $\Delta V$ is much smaller, especially in the case of Au electrodes. Second, the potential profile suggests an electronic charge transfer from the metal to the linker (somewhat larger in the case of the Ag electrodes) consistent with the electron acceptor-character of the S termination (and with the lower $\Phi$ of Ag compared to Au). A linear, but less pronounced, dependence of the potential drop on ($1/A$) is also observed for the S linker (see Fig. 4).

*Band alignments*. To understand the implications of the above results on the BA in our devices, we need to define the band edges of the chains. To this purpose, we project the density of states (PDOS) on different $CH_2$ groups, starting with the group near the contact and ending with the one in the middle of the chain. Fig. 3 shows the results of this procedure for four devices in the limit of low monolayer density

(6x6)Au-NH$_2$(CH$_2$)$_6$NH$_2$-Au(6x6) (**a**),

(6x6)Ag-NH$_2$(CH$_2$)$_6$NH$_2$-Ag(6x6) (**b**)

and

(3√3x3√3)Au-S(CH$_2$)$_6$S-Au(3√3x3√3) (**c**)

(3√3x3√3)Ag-S(CH$_2$)$_6$S-Ag(3√3x3√3) (**d**).

By examining the PDOS sequence in Fig. 3, we can distinguish the contributions originating from contact or surface states, which die out exponentially as we move away from the contacts, and the contributions from the molecular states, which become sharper and sharper as we move deeper inside the chain, allowing us to identify the band gap edges of the chain. If we denote by $E$ the energy difference between $E_F$ and the top of the molecular valence band, then from Fig. 3 we obtain $E$ = 3.7, 4.1, 3.0 and 3.5 eV, for the four cases **a**, **b**, **c**, and **d**, respectively. As one can see, the alignment is quite different in the four devices.

The same procedure was used to identify the band gap edges of the isolated monolayer in the reference systems. In this case the energy levels of the bare metal surface and of the isolated molecular monolayer were aligned using the vacuum as the reference level. Note that, for the amine terminated molecules, the reference metal surface contains adatoms (see Fig.1). These metal adatoms cause a coverage-dependent modification of $\Phi$, leading to $\Phi$ = 5.3 (4.5), 5.1 (4.3) and 5.0 (4.2) eV for the Au (Ag) surface with one adatom per (6x6), (4x4) and (3x3) unit cell, respectively. Also notice that, deep inside the molecular monolayer, the linker atoms (N or S) do not contribute to the molecular PDOS and, as a consequence, do not affect the valence band edge of the monolayer. We call $E_0$ the energy difference between $E_F$ and the top of the valence band of the isolated monolayer, see Fig. 1. Because of the large surface area and the very small values of $p_{mol}$ in all the devices, $E_0$ does not depend significantly on the linker. Indeed, using electrodes without adatoms as reference, we obtain $E_0^{Au}$=2.9 eV for the two devices with Au electrodes and $E_0^{Ag}$ = 3.7 eV for the two devices with Ag electrodes.

Similarly to the electrostatic potential, the BA in the devices depends on the monolayer density. In fact, we can correlate with the density both the variation of ΔV and that of the BA. A plot of $E - E_0$ as a function of (1/$A$), for the four different devices, is shown in Fig. 4. The effect of the monolayer coverage is very strong for the NH$_2$-terminated alkyl chains on Au electrodes, less strong when Au is replaced by Ag, and very weak for the S linked alkyl chains. The BA correlates well with the changes in ΔV inside the chain shown in Fig. 2, changes that are also shown in Fig. 4. Indeed, the value of ΔV in the middle of the chain almost perfectly matches the value of $E - E_0$ for all devices, indicating that the BA is primarily determined by the electrostatic potential inside the devices.

The sharp drop in ΔV at the contacts seen in Fig. 2 is due to the presence of a surface layer of electric dipoles. Given the almost perfect correlation between the BA and ΔV, we can write a simple equation associating the BA to the interface dipole moments (in a.u.):

$$E - E_0 = -4\pi e \frac{p_{\text{device}} - p_{\text{metal}} - p_{\text{mol}}}{A} \qquad (1)$$

In the following, we call $p_{\text{device}} - p_{\text{metal}} - p_{\text{mol}}$ the effective dipole moment $p_{\text{eff}}$. Note that Eq. (1) predicts $E = E_0$ for large $A$, i.e. in the isolated molecular limit.

Using Eq. (1), we extract from the data of Fig. 4 the values of $p_{\text{eff}}$ that are reported in Table 1. The dipole moment $p_{\text{eff}}$ can also be computed directly from:

$$p_{\text{device}} - p_{\text{metal}} - p_{\text{mol}} = e \int z \Delta n(z) dz, \qquad (2)$$

where $\Delta n(z)$ is the $(x,y)$ averaged electron density difference between a device and the corresponding reference system. A plot of $\Delta n(z)$ for several devices is shown in Fig. 5. The integral in Eq. 2 is taken over half of the unit cell shown in Fig. 5. Since $\Delta n(z)$ is symmetric with respect to the mid plane of the cell, the left and right effective contact dipole moments are the same. The effective dipoles obtained in this way are also reported in Table 1. The good agreement between these two independent evaluations of $p_{eff}$ strongly supports the validity of Eq. 1. By examining the dipole moments in Table 1, we see one major difference between the $NH_2$ and the S linked chains: in the first case, the effective dipole moment is very large, with an average value of ~4.5 D, while in the second case the effective dipole moment is very small, with an average value of ~0.2 D.

Using Eq. 1, we can explain the BA shown in Fig. 3. First, we focus on the amine linked chains. When we replace Au with Ag electrodes, $\Phi$ drops by 0.8 eV, leading to a similar increase in $E_0$. The variation of $E_0$ agrees qualitatively with the smaller value of $E$ ($E^{Au}$=3.7eV) for Au electrodes compared to the value of $E$ ($E^{Ag}$=4.1eV) for Ag electrodes reported in Fig. 3. The difference between $E^{Au}$ and $E^{Ag}$ is not exactly equal to the difference between the corresponding values of $E_0$, but the remaining difference can be explained in terms of Eq. (1), which predicts that $E^{Ag} - E^{Au} = E_0^{Ag} - E_0^{Au} - 4\pi(p_{eff}^{Ag} - p_{eff}^{Au})/A$. Using the dipole moments given in Table 1, the last term on the right hand side of the above equation accounts for the missing 0.4 eV.

In the case of the S linkers, from the right panels of Fig. 3 we extract $E^{Au}$=3.0 eV and $E^{Ag}$=3.5 eV. These values are very close to the values of $E_0^{Au}$ and $E_0^{Ag}$, as one could expect given the smaller values of $p_{eff}$ reported in Table 1 for the S linked devices. It is interesting to notice that the 0.8 eV difference between the work functions of Au and Ag electrodes leads to a change of sign of $p_{eff}$.

Based on the above discussion and the trends shown in Figs. 2, 3 and 4, we can draw the following conclusions: *(i) The BA in the long, single alkyl chain limit is independent of the link group and can be computed from the reference system alone. (ii) The convergence to the single chain limit is very slow in the case of the $NH_2$ linking groups, due to the large effective dipole moments at the contacts. Indeed even for a 6×6 supercell, the BA is off by more than 0.5 eV from the isolated molecular limit. (iii) The convergence to the single chain limit is more rapid in the case of the S anchoring groups due to very small effective dipole moments at the contacts. In this case the single chain limit is already achieved in practice with a $3\sqrt{3}\times3\sqrt{3}$ supercell.*

At this point a brief comment on the effect of the DFT approximations on the BA is in order. In exact DFT the highest occupied Kohn-Sham eigenvalue should be equal to the ionization potential, but in practical calculations the highest occupied Kohn-Sham level depends on the adopted functional approximation. The DFT error on metal work functions is small, as shown, for instance, by the good agreement of our calculated values for Au and Ag with experiment. Instead, the error on the ionization potential of an alkane chain is quite large. In fact the top valence band edge of our reference molecular chain (see Fig.1) is located at ~8 eV below the vacuum level, while experimentally the same level should be at ~ 11 eV below vacuum[30], indicating that we underestimate $E_0$ by ~ 3 eV in our calculations. This error originates mostly from the self-interaction error of approximate DFT functionals. On the other hand, our analysis shows that the change from $E_0$ to $E$ is entirely due to the charge rearrangement following the formation of the contacts between the molecule and the electrodes. This charge rearrangement should be accurately described by approximate functionals like the one adopted here. How the alignment errors affect the calculated conductance of molecular devices deserves a thorough separate investigation.

*Discussion.* We identify several interesting features in the electron density difference Δ*n(z)* in Fig. 5. In the case of the NH$_2$ anchoring group there is a significant transfer of electronic charge from NH$_2$ to the Au electrodes. A weaker, but still important electron transfer occurs when Au is replaced by an Ag electrode. In the case of the S anchoring group the opposite behavior occurs, namely some electronic charge is transferred from the electrodes to the anchoring groups. The electron transfer is stronger for the Ag electrodes, where it is large enough to change the sign of $p_{\text{eff}}$. The electron transfer between the linkers and the metal electrodes is consistent with the known donor/acceptor character of the amine and sulfur groups.

A striking feature of Δ*n(z)* in Fig. 5 is that, regardless of the direction of the electron flow between electrodes and anchoring groups, the alkane chain always looses electrons. Equally interesting is the fact that Δ*n(z)* never changes sign inside the chain, in spite of the fact that it undergoes a significant variation. These features of Δ*n(z)* can be understood in terms of the asymptotic expression of Δ*n(z)* away from the contacts, which takes the form:[31][32]

$$\Delta n(z) \to -\rho_0 \frac{dE_k}{dq}\bigg|_{E_k=E_F} \frac{u_{k_F}(z)^2}{u_{k_F}(z_0)^2} \frac{e^{2ik_F(z-z_0)}}{z-z_0}. \qquad (3)$$

Here $z_0$ is the midpoint between the anchoring group and the first C atom of the alkane chain, *k* is the wavevector of the Bloch functions of the (infinite) periodic molecular chain, $q = \text{Im}[k]$, $u_{k_F}(z)$ is the (*x*,*y*) average of the periodic part of the relevant evanescent Bloch function of the alkane chain at $E_F$, and $\rho_0$ is the local density of states near $z_0$, evaluated at $E_F$. Since $E_F$ falls inside the gap of the chain, $k_F$ has a strictly positive imaginary part, which leads to the exponential decay of Δ*n(z)* inside the chain, with an exponent equal to $\beta = 2\text{Im}[k_F]$. The exponential decay of Δ*n(z)* is visible in all the plots in Fig. 5, but is more clearly visible in the case of the longer chains. In alkane chains, the relevant complex band energies are real,[33] and thus $u_k(z)$ is a real valued function. Since $u_k(z)$ appears at the second power in Eq. 3, we understand why $\Delta n(z)$ never changes sign as $z$ is varied.

Furthermore, at the branch point, defined as the turning point of the complex band, $\frac{dE_k}{dq} = 0$, and the right hand side of Eq. 3 vanishes. At this point, a global change of sign of $\Delta n(z)$ takes place. More precisely, if $E_F$ is below the branch point, $\frac{dE_k}{dq} > 0$ and the right hand side of Eq. 3 is negative, while if $E_F$ is above the branch point, $\frac{dE_k}{dq} < 0$ and the right hand side of Eq. 3 is positive. In the single chain limit, $E \to E_0$, and $E_0$ is far below the branch point of the alkane chain. This observation explains why $\Delta n(z)$ inside the chain is negative in all devices, regardless of the sign of the electron transfer between the linker group and the metal electrode.

In the asymptotic expression, Eq. (3), only $\rho_0$ depends on the chemistry of the contact. All the remaining factors depend solely on the complex band structure of the alkane chain and on the position of $E_F$ relative to the band edges of the chain. Based on this observation, we can identify the major difference between amine and thiol linked chains. The BA in the isolated limit of a long molecular chain is very similar in the two cases, meaning that $\beta$ and $u_k(z)$ are essentially the same in the two cases. Yet, in Fig. 5, one sees a big difference in the evanescent fraction of $\Delta n(z)$, when comparing amine and thiol linking groups. *This indicates that the difference between the two cases originates from a large difference in the value of $\rho_0$, which is large for amine linked chains and very small for thiol linked chains.* The origin of this difference should be traced to the way in which the linker group binds to the metal. As we have already pointed out, the charge transfer between linker groups and molecular chains is very small. Thus, the charge transfer in Fig. 5 is mainly determined by the chemistry between the electrode and the linker group. The evanescent $\Delta n(z)$ is an aftermath of this process, which attributes a large value to $\rho_0$ in the case of the $NH_2$ linker and a lower value to it in the case of the S linker.

The following mechanism emerges for the BA: in a first step, a charge transfer is initiated by the chemistry between the linker and the metal, leading to a chemical contact dipole and to a definite value of $\rho_0$. An evanescent $\Delta n(z)$ sets in inside the chain, further contributing to the effective dipole of the device. The dipole moment that originates from the evanescent wave contribution to $\Delta n(z)$ can be calculated from Eq. 3. This dipole depends on the BA $E$. For example, a large evanescent dipole occurs when $E_F$ is close to the valence band and a small one occurs when $E_F$ is near the branch point. This dependence, together with Eq. 1, ultimately gives rise to a self-consistent equation that determines the level alignment in the devices. In this equation, the only inputs are the chemical dipole and the value of $\rho_0$. A quantitative analysis of this mechanism will be presented in future work.

*Implications.* The former conclusions have important implications for both experiment and theory. They explain the very weak dependence of β on the linker groups found experimentally.[22, 24] They also suggest a simple rule to predict the changes in the BA when different metals are used for the electrodes. For example, in the isolated chain limit, one should expect a difference of 0.8 eV in the BA when Au is replaced by Ag in the electrodes. This difference, however, has only a minor effect on the β coefficient of the alkane chain because the relevant complex band is flat within a large energy window inside the gap of the chain.[34] This observation is in agreement with previous experimental findings.[7] Finally, our study also indicates that the BA may depend strongly on the monolayer coverage, suggesting that an alternative route for controlling the transport characteristics of monolayer devices is to control the coverage.

**Acknowledgement.** This work was supported by NSF Grant DMR-0213706 to the MRSEC-Princeton Center for Complex Materials. RC acknowledges also partial support from DOE grant DE-FG02-05ER46201. EP acknowledges support from Yeshiva University.

**Supporting Information Available:** Computational details. This material is available free of charge via the Internet at http://www.pubs.acs.org.

**Table 1.** Charge transfers (in units of electron charge) and effective dipoles (in Debye) calculated with different super cells. Charge transfers are determined by integrating $\Delta n(z)$ over the molecule from the left to the right contact, each taken at the midpoint between the metal and the linker atom. A negative charge transfer means that the molecule has lost electrons to the metal electrodes.

| Metal-$NH_2(CH_2)_nNH_2$-Metal | Charge transfer (e) | Dipole Moment Eq. (2) (Debye) | Dipole Moment Eq. (1) (Debye) | Metal-$S(CH_2)_nS$-Metal | Charge transfer (e) | Dipole Moment Eq. (2) (Debye) | Dipole Moment Eq. (1) (Debye) |
|---|---|---|---|---|---|---|---|
| Au(3x3), n=6 | -0.41 | 4.08 | 3.84 | Au($2\sqrt{3}\times2\sqrt{3}$), n=6 | -0.05 | 0.25 | 0.33 |
| Au(4x4), n=6 | -0.51 | 4.32 | 4.70 | Au($3\sqrt{3}\times3\sqrt{3}$), n=6 | -0.03 | 0.06 | 0.18 |
| Au(6x6), n=6 | -0.56 | 5.04 | 5.62 | Au($2\sqrt{3}\times2\sqrt{3}$), n=16 | -0.02 | 0.19 | |
| Au(4x4), n=16 | -0.45 | 3.41 | | | | | |
| Ag(4x4), n=6 | -0.36 | 4.03 | 3.50 | Ag($2\sqrt{3}\times2\sqrt{3}$), n=6 | +0.20 | -1.06 | -1.25 |
| Ag(6x6), n=6 | -0.33 | 4.13 | 3.55 | Ag($3\sqrt{3}\times3\sqrt{3}$), n=6 | +0.21 | -0.82 | -1.01 |

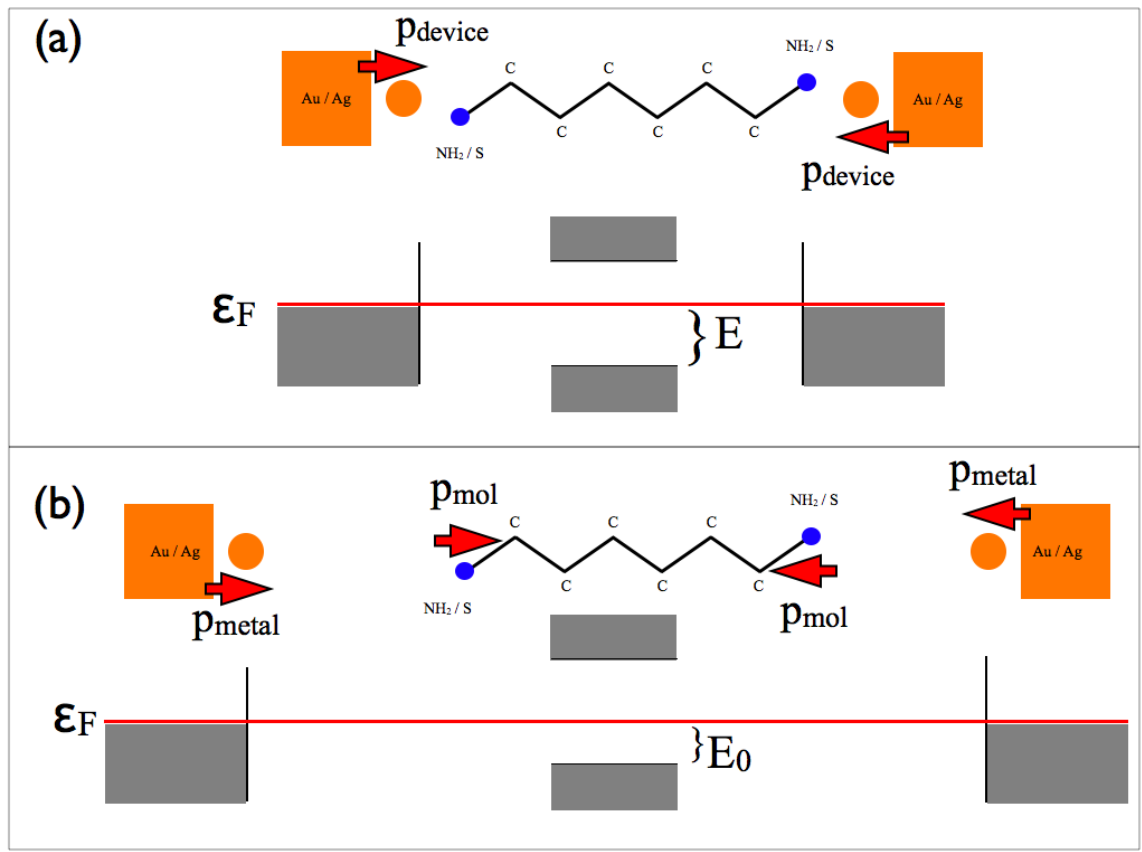

**Figure 1.** Schematic representation of: (a) a molecular device, with the corresponding band alignment, $E$, and dipole moment at the device contacts, $p_{device}$; (b) the reference system of non-interacting molecule + electrodes, with the corresponding band alignment, $E_0$, and molecule and electrode dipole moments, $p_{mol}$ and $p_{metal}$. The band alignment is equal to the energy difference between $E_F$ and the HOMO of the alkyl chain.

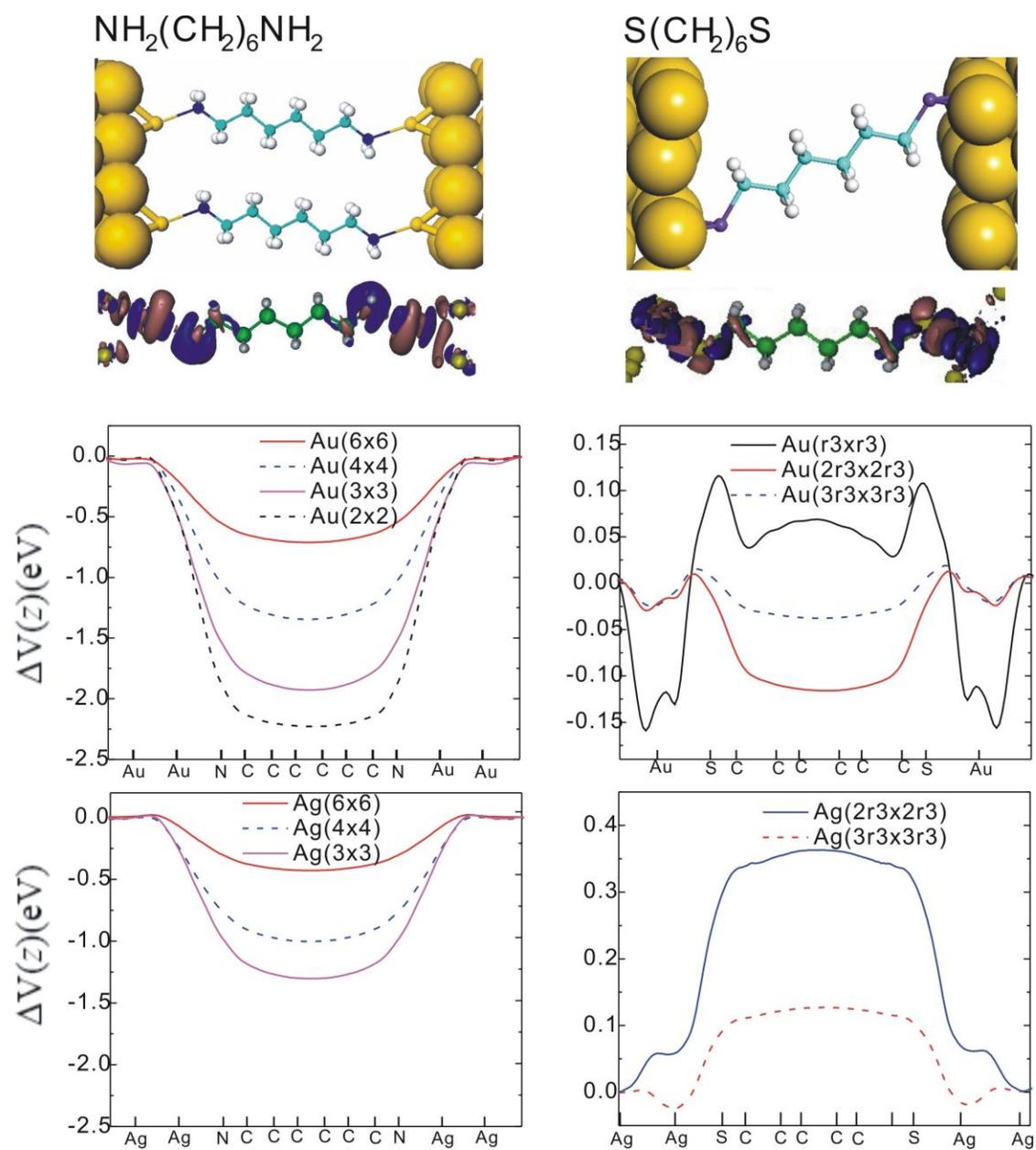

**Figure 2.** From top to bottom: geometry, charge density difference $\Delta n(\mathbf{r})$ (isosurfaces), and electrostatic potential energy difference $\Delta V(z)$ with Au (above) and Ag (below) electrodes. Left: M-NH$_2$ C$_6$H$_{12}$ NH$_2$-M junctions with amine-terminated molecular chains; right: M-SC$_6$H$_{12}$ S-M junctions with thiolate-terminated molecular chains.

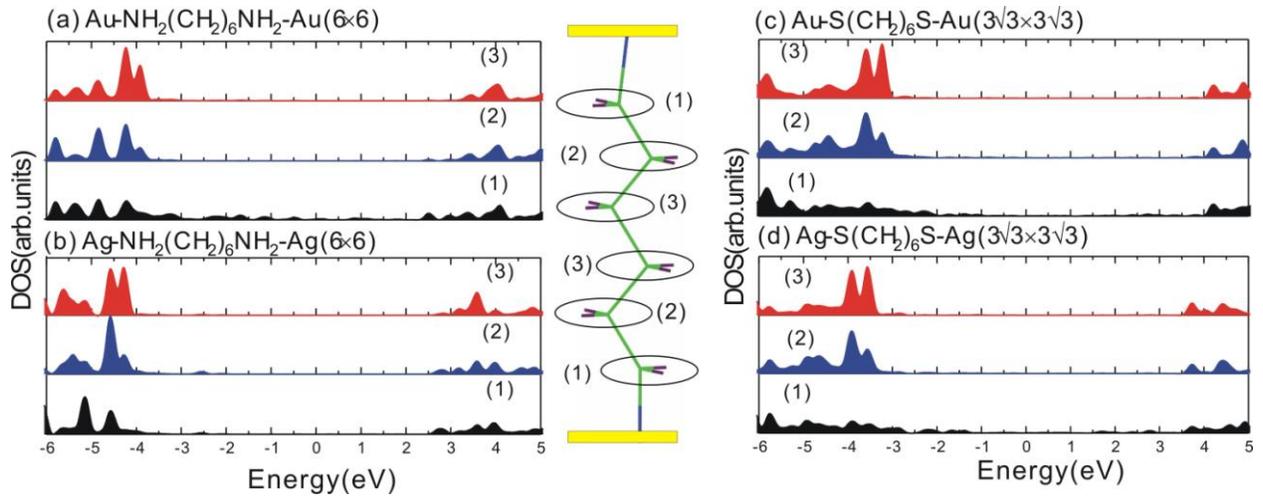

**Figure 3.** Projected densities of states onto the different $CH_2$ groups along the molecular chain, as indicated in the middle panel, for the following devices: (a) Au-$NH_2C_6H_{12}NH_2$-Au, 6×6 supercell; (b) Ag-$NH_2C_6H_{12}NH_2$-Ag, 6×6 supercell; (c) Au-$SC_6H_{12}S$-Au, $3\sqrt{3}\times3\sqrt{3}$ supercell; (d) Ag-$SC_6H_{12}S$-Ag, $3\sqrt{3}\times3\sqrt{3}$ supercell. The energy zero is set at $E_F$.

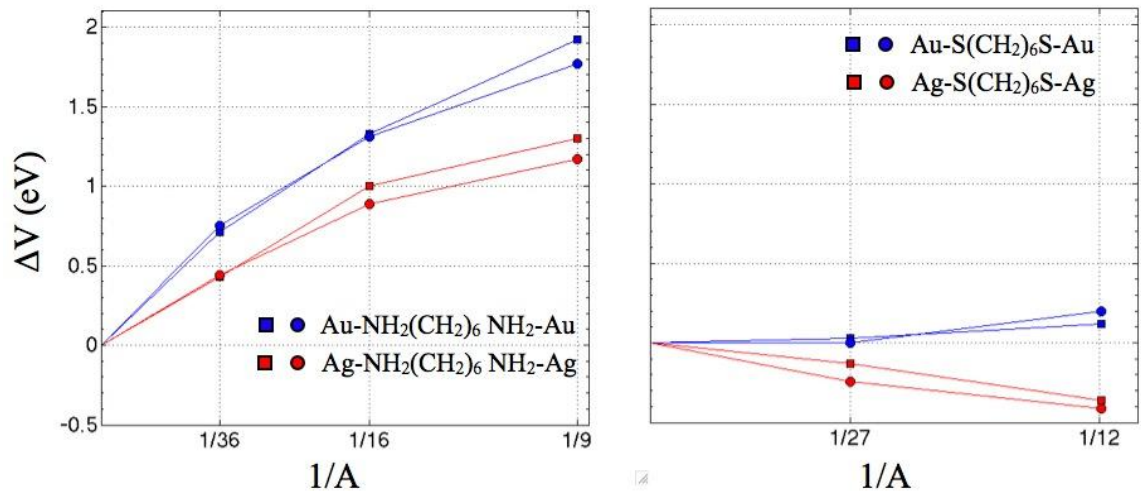

**Fig. 4** Dependence of $E$-$E_0$ (circles) and $\Delta V$ (squares) on the inverse size $1/A$ of the supercell. The values of $\Delta V$ are derived from the plots in Fig. 2, those of $E$-$E_0$ are computed as explained in the text. Left: amine-terminated molecular chains; right: thiolate-terminated chains.

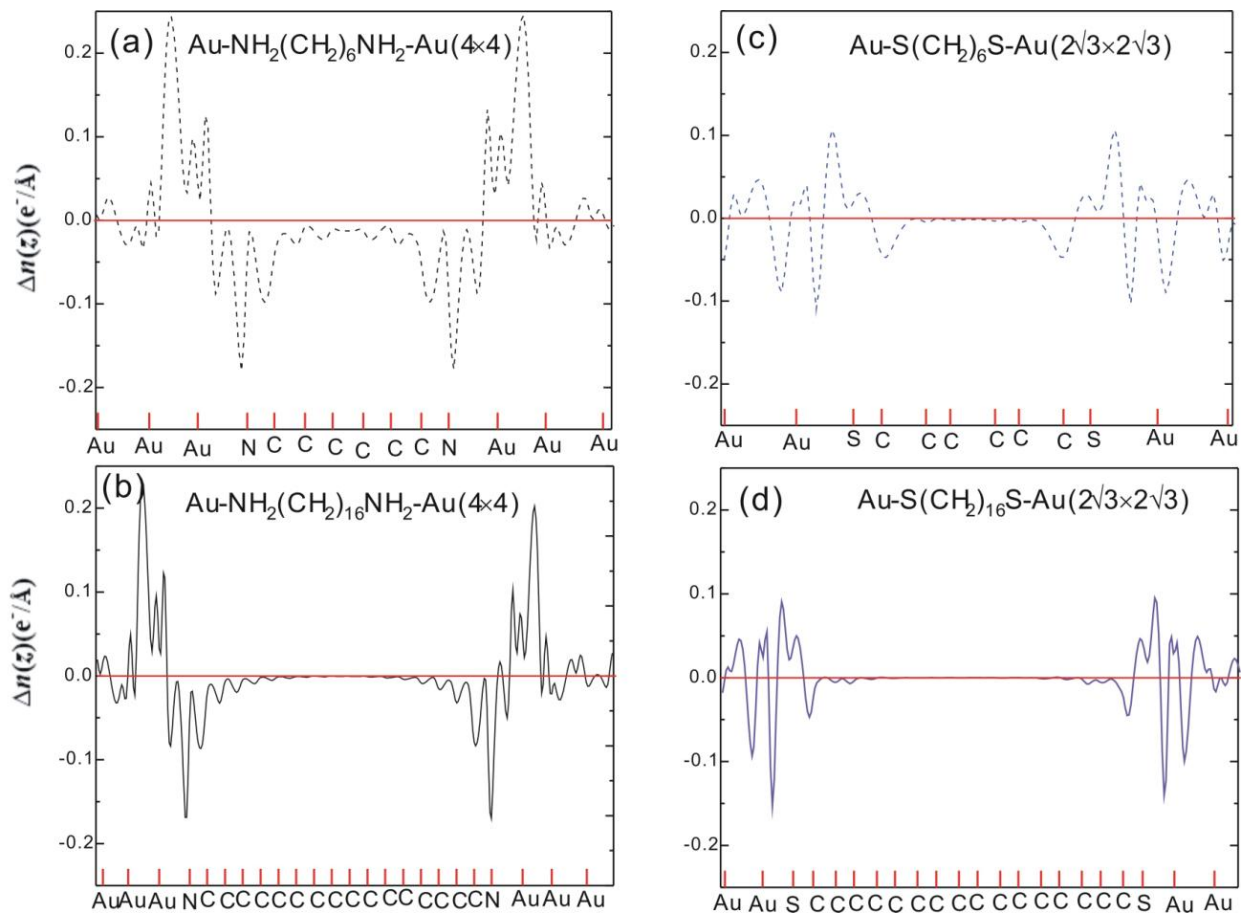

**Fig. 5** Charge density difference $\Delta n$ ($z$) for : (a) Au-NH$_2$C$_6$H$_{12}$NH$_2$-Au, 4×4 supercell; (b) Au-NH$_2$C$_{16}$H$_{32}$NH$_2$-Au, 4×4 supercell; (c) Au-SC$_6$H$_{12}$S-Au, 2√3 ×2√3 supercell; (d) Au-SC$_{16}$H$_{12}$S-Au, 2√3×2√3 supercell. The positions of the atoms along the molecular junctions are indicated.